\documentstyle[psfig,twocolumn,prl,aps]{revtex}
\newcommand{\NP}[1]{ Nucl.\ Phys.\ {\bf #1}}
\newcommand{\ZP}[1]{ Z.\ Phys.\ {\bf #1}}

\newcommand{\PL}[1]{ Phys.\ Lett.\ {\bf #1}}

\newcommand{\PR}[1]{Phys.\ Rev.\ {\bf #1}}
\newcommand{\PRL}[1]{ Phys.\ Rev.\ Lett.\ {\bf #1}}

\begin{document}

\twocolumn[\hsize\textwidth\columnwidth\hsize\csname
@twocolumnfalse\endcsname


\title{Testing QCD with Hypothetical Tau Leptons$^1$}

\author{
S. J. Brodsky, J. R. Pel\'aez$^2$ and N. Toumbas\\
{\small\em Stanford Linear Accelerator Center}\\
{\small\em Stanford University,
Stanford, California 94309. U.S.A.}}

\maketitle

\begin{abstract}
We construct new tests of  perturbative QCD by considering  a hypothetical
$\tau$ lepton of arbitrary mass, which decays hadronically through the
electromagnetic current.  We can explicitly compute its hadronic
width ratio directly as an integral over the $e^+ e^-$ annihilation
cross section ratio, $R_{e^+e^-}$.   Furthermore, we can design a set of
commensurate scale relations and perturbative QCD tests by varying the
weight function away from the form associated with the $V-A$ decay
of the physical
$\tau$.  This method allows the wide range of the $R_{e^+e^-}$ data to be
used as a probe of  perturbative QCD.
\end{abstract}
]
\footnotetext[1]{Research partially supported
by the U.S. Department of Energy under contract DE-AC03-76SF00515
and the Spanish CICYT under contract AEN93-0776.}
\footnotetext[2]{E-mail:pelaez@slac.stanford.edu. On
leave of absence from the Departamento de F\'{\i}sica Te\'orica.
Universidad Complutense. 28040 Madrid. Spain.}

The hadronic width of the $\tau$ lepton is potentially one of the most
important sources for the  high precision determination of the coupling
$\alpha_{\overline{MS}}$ of QCD \cite{Braaten,Groote}.  The
perturbative QCD (PQCD) analysis of the $\tau$ width has been refined
by constructing moments of hadronic decay distributions which 
minimize sensitivity to the low energy part of the hadronic
spectrum \cite{Davier}. 
 However, it is still  uncertain
whether the $\tau$ mass is sufficiently high to trust PQCD, 
particularly due to the strong distortion of hadronic
final state interactions \cite{Smilga}.

In this paper we construct new renormalization scheme-independent
tests of  PQCD which we can apply, not only to the physical
$\tau$ lepton, but also to a  hypothetical
$\tau$ lepton of arbitrary mass which decays hadronically through the
vector current. Such hypothetical $\tau$ leptons, with masses $M<M_\tau$,
have already been considered in ref.\cite{Davier}.
 We can obtain empirical values for the
hypothetical lepton's hadronic width and moments directly as integrals
over the measured 
$R_{e^+e^-}=\sigma(e^+e^-\rightarrow {\rm hadrons})/
\sigma(e^+e^-\rightarrow\mu^+\mu^-)$.  
 As we shall show, these tests are
fundamental properties of QCD which can serve as necessary conditions for
the applicability of perturbation theory.

Quantum field theoretic predictions  which relate physical
observables cannot depend on theoretical
conventions such as the choice of renormalization scheme or
scale.  The most well-known  example
is the ``generalized Crewther
relation" \cite{Kataev} in which the leading twist 
PQCD  corrections to the Bjorken sum rule at a given
lepton momentum transfer
$Q^2$ are inverse to the QCD corrections to $R_{e^+e^-}$ at
a corresponding CM energy squared,
$s^* = s^*(Q^2)$,  independent of
renormalization scheme. The ratio of the scales
$s^*/Q^2$ has been computed to NLO in PQCD.
Such leading-twist predictions between observables are
called ``commensurate scale relations" and are identical for
conformal and nonconformal theories \cite{CSR}.

Another important example is the commensurate scale relation 
between the PQCD correction to the $\tau$
lepton's width ratio,
$R_{\tau} = \Gamma(\tau^-\rightarrow{\nu_{\tau} +
{\rm hadrons}})/\Gamma(\tau^-\rightarrow{\nu_{\tau}e^-\bar{\nu_e}})$,
and those to $R_{e^+e^-}$.
Assuming for now $f$ massless flavors, PQCD yields
\begin{equation}
R_{e^+e^-}(\sqrt{s})=(3\sum_f q_f^2)\left[1+\frac{\alpha_R(\sqrt{s})}{\pi}\right],
\label{ReePQCD}
\end{equation}
where $\alpha_R$ can be written as a series
in
$\alpha_s/\pi$ in some renormalization scheme.
Note that $\alpha_R$ is an effective charge
\cite{effcha} because it
satisfies the Gell-Mann-Low renormalization group equation with the same
coefficients $\beta_0$ and $\beta_1$ as the usual coupling
$\alpha_s$ (differing only through the third and higher
coefficients of the $\beta$-function). Similarly we can define an effective
charge $\alpha_\tau$ as follows
\begin{equation}
R_{\tau}(M_\tau)=R^0_{\tau}(M_\tau)
\left[1+\frac{\alpha_\tau(M_\tau)}{\pi}\right].
\label{RtauPQCD}
\end{equation}
Leading-twist QCD predicts
\begin{equation}
\alpha_\tau(M_\tau) = \alpha_R(\sqrt{s^*})
\label{CSRtau}
\end{equation}
to all orders in perturbation theory. The
ratio of the commensurate scales is known in
NLO PQCD:
\begin{equation}
\frac{\sqrt{s^*}}{M_\tau} = \exp\left[ -{19\over 24} - {169\over 128}
{\alpha_R(M_\tau)\over \pi} + \cdots\right].
\label{CSStau}
\end{equation}
This result was originally obtained in \cite{CSR} by using NNLO
predictions for
$\alpha_R$ and $\alpha_\tau$ obtained in the ${\overline{MS}}$
scheme and eliminating
$\alpha_{\overline{MS}}$.  However, as we shall show here, the
QCD prediction for
$\sqrt{s^*}/M_\tau$ also follows from the fact that both effective
charges evolve with universal
$\beta_0$ and
$\beta_1$ coefficients.   
The fact that $R_\tau$ can be expressed as
\footnote{We have used $\vert{V_{ud}}\vert^2 + \vert{V_{us}}\vert^2 = 1$,
as in \cite{taudec}. Note that in order to
include NNLO
corrections in $\alpha_{\tau}$, we must modify the $O(\alpha_s^3)$
coefficient of $\alpha_R$ by setting $(\sum_f{q_f})^2=0$.} 
\begin{eqnarray}
R_{\tau}(M_\tau) &=&
\frac{2}{(\sum_f{q_f^2})}\label{defRt} \\ \nonumber
&\times&\int^{M_{\tau}^2}_0\frac{\,d\,s}{M_{\tau}^2}
\left(1 -
\frac{s}{M_{\tau}^2}\right)^2\left(1 +
\frac{2s}{M_{\tau}^2}\right)R_{e^+e^-}(\sqrt{s})
\end{eqnarray}
implies, by the mean value theorem, that 
$\alpha_R$ and $\alpha_\tau$ are related by a scale shift.
However, the prediction for the ratio
$\sqrt{s^*}/M_\tau$ in eq.(\ref{CSStau}) is a specific property of PQCD.

A definitive empirical test of the commensurate relation, eq.(\ref{CSStau}),
is problematic since there is only one $\tau$ lepton 
in nature, and its mass  seems uncomfortably low for tests of
leading-twist QCD.
However, we can construct new tests of PQCD by considering
a hypothetical
$\tau$ lepton of arbitrary mass $M$ which decays hadronically through the
vector current.  Then
we can explicitly compute its hadronic
width ratio as an integral over the measured $R_{e^+e^-}$.  Furthermore,
we can design a set of commensurate scale relations and PQCD tests by
varying the weight function away from the form associated with the $V-A$
decays of the physical
$\tau.$  Thus we can use the full range of the $R_{e^+e^-}$ as a 
novel test of PQCD.
As we shall show, such a test
must also take into account specific effects attributable to
the $s
\bar s, c\bar c, b \bar b$ quark thresholds.  Also, following \cite{PQW}
we  shall smear the annihilation data in energy
in order to eliminate resonances and other distortions of final
state interactions.  By smearing $R_{e^+e^-}$ over a range of energy,
$\Delta E$, we focus the physics to the time $\Delta t = 1/\Delta
E $ where an analysis in terms of PQCD quark and gluon
subprocesses is appropriate. Therefore, this method
can also be interpreted as an additional test of duality.
Scheme-independent relations between $R_{e^+e^-}$ and
$\tau$ decay have also been recently discussed in \cite{Groote}.

Given $\alpha_R$, we  can construct
new effective charges as follows:
\begin{equation}
\alpha_f(M) \equiv \frac{\int^{M^2}_0 \frac{d\,s}{M^2}\,f\left(
\frac{s}{M^2}\right)\alpha_R(\sqrt{s})}{\int^{M^2}_0 \frac{d\,s}{M^2}f\left(
\frac{s}{M^2}\right)}.
\label{alphaRf}
\end{equation}
We can choose $f(x)$ to be any smooth, integrable function of
$x=s/M^2$.
( For the particular choice, $f_\tau(x) = \left(1 - x\right)^2
\left(1 + 2x\right)$,
$\alpha_R^f$ is simply $\alpha_{\tau}$.)
The mean value theorem then implies
\begin{equation}
\alpha_f(M)=\alpha_R(\sqrt{s^*_f}),\hspace{2cm}    0\leq s^*_f\leq M^2.
\label{MVT}
\end{equation}
Dimensional analysis ensures that $\sqrt{s^*_f}=\lambda_f\, M$, where
$\lambda_f$ possibly depends on $\alpha_R$. To obtain an estimate for
$\lambda_f$ we consider the running of $\alpha_R$ up to third
order:
\begin{eqnarray}
&&\frac{\alpha_R(\sqrt{s})}{\pi}= \frac{\alpha_R(M)}{\pi} -
\frac{\beta_0}{4} \ln{\left(\frac{s}{M^2}\right)}
\left(\frac{\alpha_R(M)}{\pi}\right)^2  + \label{aR2order} \\
\nonumber
&&\quad + \frac{1}{16}\left[
\beta_0^2 \ln^2\left(\frac{s}{M^2}\right)-
\beta_1 \ln\left(\frac{s}{M^2}\right)\right]
\left(\frac{\alpha_R(M)}{\pi}\right)^3
\ldots
\end{eqnarray}
We substitute for $\alpha_R$
in eq.(\ref{alphaRf}) to find
\begin{eqnarray}
\frac{\alpha_f(M)}{\pi} &=& \frac{\alpha_R(M)}{\pi} -
\frac{\beta_0}{4}\left(\frac{I_1}{I_0}\right)
\left(\frac{\alpha_R(M)}{\pi}\right)^2 \nonumber \\
&+& \frac{1}{16}\left[\beta_0^2 \left(\frac{I_2}{I_0}\right)-
\beta_1 \left(\frac{I_1}{I_0}\right)\right]
\left(\frac{\alpha_R(M)}{\pi}\right)^3
\ldots,
\label{aRf2order}
\end{eqnarray}
where  $I_l = \int^{1}_0f(x)(\ln{x})^ld\,x$.
By setting $s=s^*$ in eq. (\ref{aR2order}) and comparing with
eq. (\ref{aRf2order}), we extract
\begin{equation}
\lambda_f = \exp{\left[\frac{I_1}{2I_0} +
\frac{\beta_0}{8}\left(\left(\frac{I_1}{I_0}\right)^2 - \frac{I_2}{I_0} 
\right)\frac{\alpha_R(M)}{\pi}\right]}.
\label{sstar}
\end{equation}
Note that if $f(x)$ is positive on the interval $[0,1]$, then
$I_1/I_0$ is negative as expected. Using $f_\tau(x)$, eq. (\ref{sstar})
is nothing but eq. (\ref{CSStau}). Also, since $\lambda_f$ is a
constant to leading order, $\alpha_f$ should satisfy the same RG
equation as $\alpha_R$ with the same coefficients $\beta_0$ and
$\beta_1$. In other words, $\alpha_f$ is an effective
charge.

We can now  study integrals over
$R_{e^+e^-}$ data with different weight functions
$f(x)$ and vary $M$ to see whether we
obtain the PQCD behavior. In general,
the weight function $f(x)$ should be
chosen to suppress the low energy region,
where non-perturbative effects are important.
Thus, in the following, we will set $f(x)=x^k$,
where $k$ is some positive number. In such a case, we have that
\begin{equation}
\alpha_k(M) = \alpha_R(\lambda_k\,M)\quad \hbox{with}\quad
\lambda_k=e^{I_{1k}/2I_{0k}},
\label{atk}
\end{equation}
where $I_{1k} = \int^{1}_0x^k\ln{x}\,d\,x$ and $I_{0k} =
\int^{1}_0x^k\,d\,x$.
Note
that as k increases, $I_{1k}/2I_{0k}\rightarrow 0$, and, therefore,
$\sqrt{s^*}\rightarrow{M}$.
For very large k, we lose sensitivity to the details of PQCD.
It is particularly interesting to use
such a test to probe the energy region close to
the $\tau$ mass $M_{\tau}$.

The main difficulty in comparing with $R_{e^+e^-}$ data
is that we can no longer consider massless flavors and that
we observe hadrons instead of quarks.

Following \cite{PQW} the effect of quark masses can be
approximately taken into account  if we use:
\begin{eqnarray}
R_{e^+e^-}(\sqrt{s})&=&3\sum_1^f q_i^2 \frac{v_i(3-v_i^2)}{2}\left[ 1+g(v_i)
\frac{\alpha_R(\sqrt{s})}{\pi}\right]\nonumber \\
&\equiv&R_0(\sqrt{s})+R_{Sch}(\sqrt{s})\frac{\alpha_R(\sqrt{s})}{\pi}\label{Rsch} \\
g(v)&=&\frac{4\pi}{3}\left[\frac{\pi}{2v}-\frac{3+v}{4}\left(
\frac{\pi}{2}-\frac{3}{4\pi}\right)\right]\label{g(v)}
\end{eqnarray}
where $v_i=\sqrt{1-4m_i^2/s}$ is the velocity of the initial
quarks in their CM frame. The $v_i(3-v_i^2)/2$ factor is the parton model
mass dependence and $g(v)$ is a QCD modification \cite{QCDSch} of
the Schwinger positronium corrections \cite{Sch}. In principle, all
these corrections spoil the relation in eq.({\ref{atk}}).
However these factors are
unity for energies well above their corresponding thresholds.

Nevertheless, we still cannot compare directly with
the data since there
is no direct correspondence between quark and hadronic
thresholds. To obtain a meaningful comparison
we have to smear both the PQCD results and the data.
Following \cite{PQW} we define smeared quantities as follows:
\begin{equation}
\bar{R}(\sqrt{s})=
\frac{\Delta}{\pi}\int^\infty_0 \frac{R(\sqrt{s'})}
{(s-s')^2-\Delta^2}\,d\,s'
\label{smear}
\end{equation}
Note that in the $\Delta\rightarrow0$ limit, we recover the original quantity.
In what follows we use the standard value $\Delta=3\,\hbox{GeV}^2$
\cite{PQW,MaSt}. The smearing effect can be seen by comparing Fig.1,
which shows an interpolation of the $R_{e^+e^-}$ data,
\cite{DATA}, with Fig.2.
For completeness, we also include in Fig.2 the smeared results from
NLO PQCD and from the naive parton model ($\alpha_R=0$).
\begin{figure}
\hbox{
\psfig{file=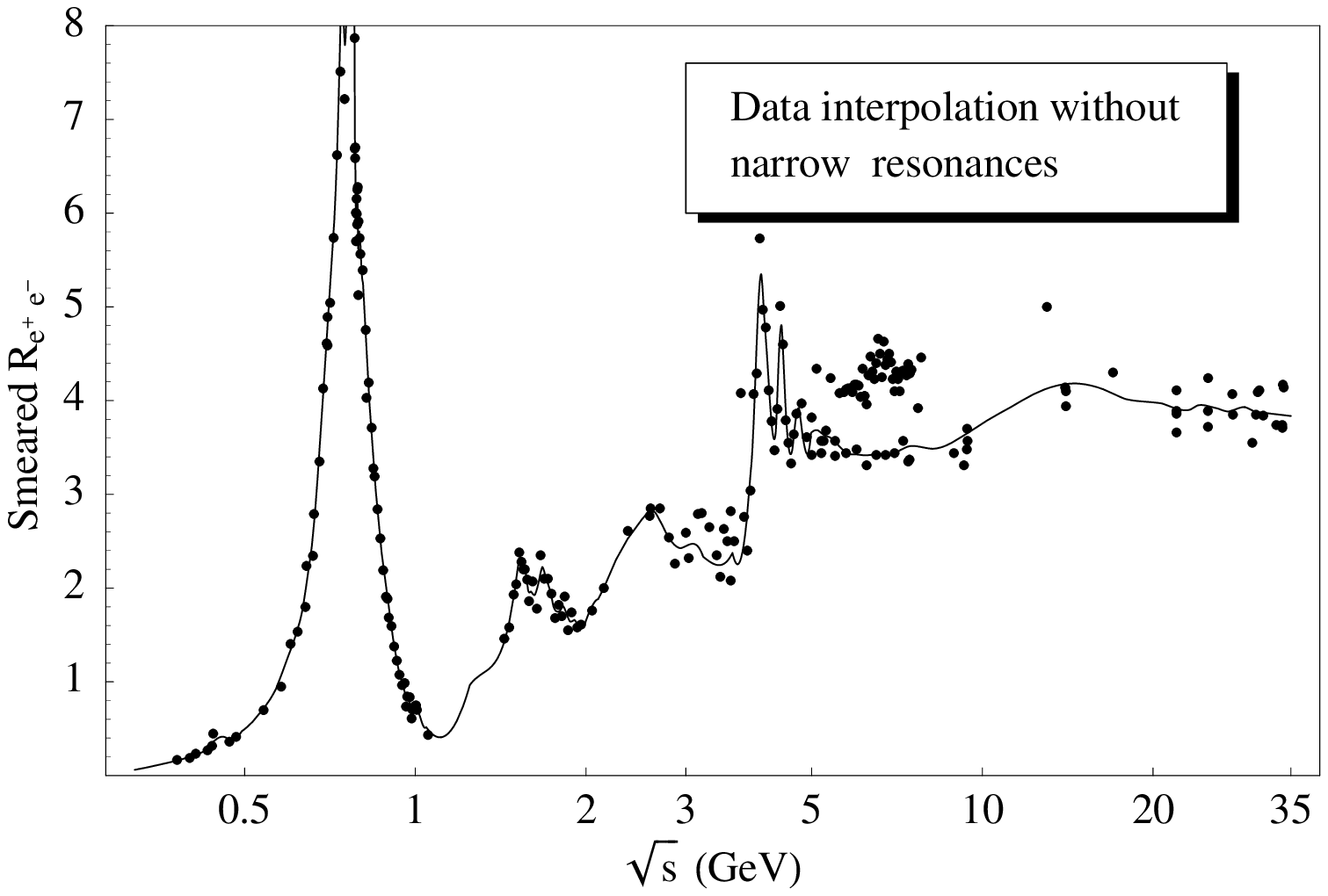,width=8.5cm}
}
{\footnotesize {\bf Figure 1:}
Interpolation of the central values of $R_{e^+e^-}$ data \cite{DATA}. Narrow
resonances are taken into account using their Breit-Wigner form.
Note that there seems to be a
discrepancy in the central values of experiments
 between 5 and 10 GeV, that
above 20 GeV  we have two or 
three clearly different central values at the same $\sqrt{s}$,
and that the point at
13 GeV is much higher than other nearby data.}
\end{figure}

\begin{figure}
\hbox{
\psfig{file=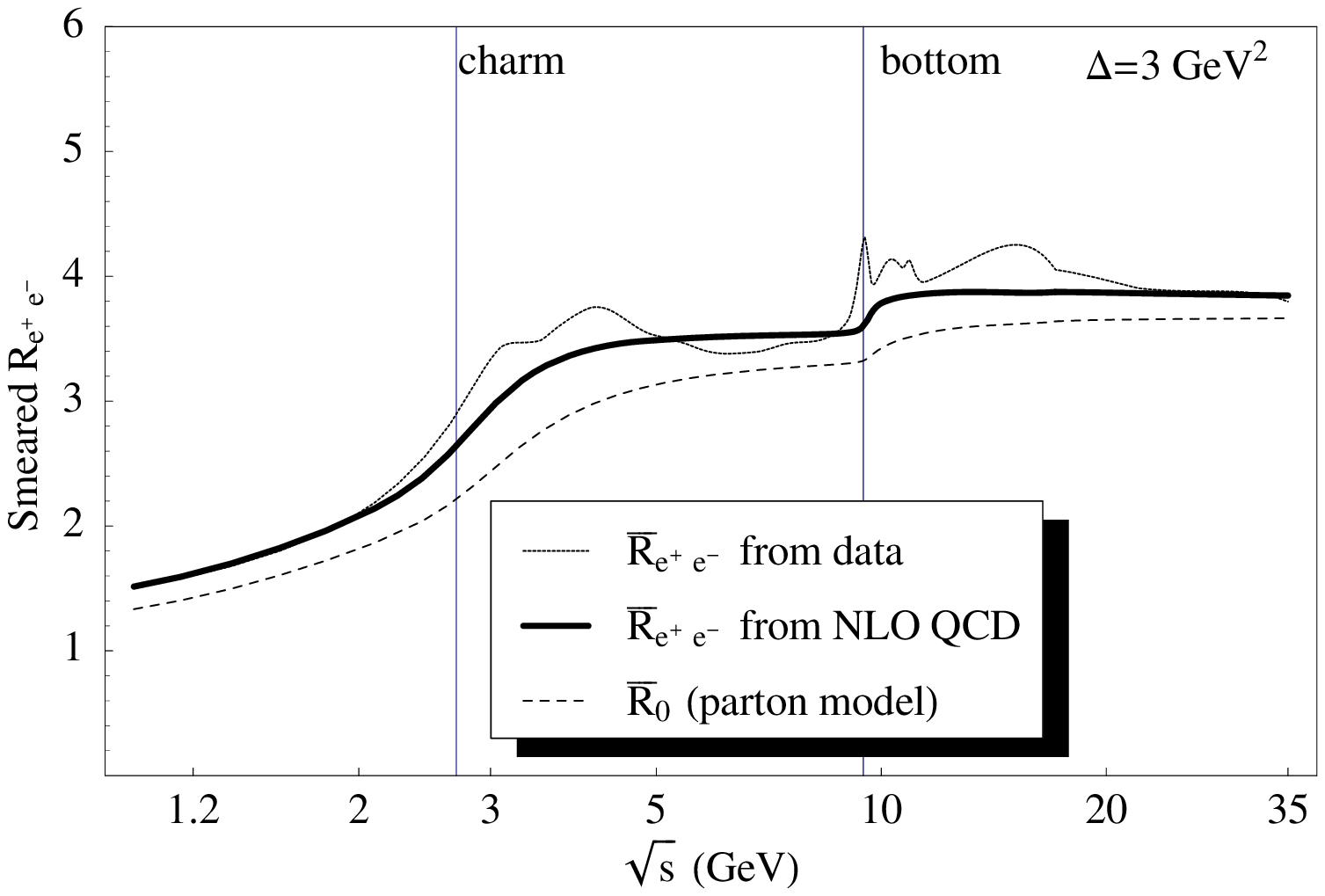,width=8.5cm}
}
{\footnotesize {\bf Figure 2:}
Effect of smearing on $R_{e^+e^-}$.}
\end{figure}
In order to integrate over $R_{e^+e^-}$, we 
need to interpolate, {\em but not fit}, the data. Note that
{\em any fit using the QCD functional dependence will  always satisfy
the commensurate scale relations}, even if its quality
is poor. To avoid this bias, we have interpolated the 
central values of the data
 by means of ``$r$-term simple moving averages''
 up to $30$ GeV (to avoid electroweak contributions). That is,
if we have a series of raw data ${z_1,...z_n}$, we obtain the new
set of smoothed data $\sum_{j=0}^{r-1}w_j\,z_{t-j}$ for $t=r,...n$,
with $\sum_{j=0}^{r-1}w_j=1$.
We have used $r$ ranging from 2 to 6 for different energy regions and
our moving averages are ``simple'' because all the weights $w_j$
are equal. Finally, the resulting smoothed data have been interpolated
using cubic-splines.
In addition, the narrow resonances that do not
appear in Fig.1
are implemented using the Breit-Wigner formula.

We have thus eliminated the QCD biases
up to $30$ GeV. Above that energy we have matched a
logarithmic function whose functional dependence is inspired by QCD, but
its contribution in the smearing integrals is negligible for
small $\sqrt{s}$.

Unfortunately, we cannot extract directly the effective
charges from their corresponding smeared ratios since
they are multiplied by other functions inside the smearing integral.
However, using eqs.(\ref{Rsch}) and (\ref{smear}), we define
smeared charges:
\begin{equation}
\bar{\alpha}_R(\sqrt{s})=
\frac{\bar{R}_{e^+e^-}(\sqrt{s})-\bar{R}_0(\sqrt{s})}{\bar{R}_{Sch}(\sqrt{s})},
\end{equation}
and similarly for $\bar{\alpha}_k$.
In the massless $\Delta\rightarrow0$ limit we recover
the standard effective charges.
We expect the smeared charges to satisfy
eq.(\ref{sstar}) in energy regions where the threshold corrections
can be neglected. 

\begin{figure}
\hbox{
\psfig{file=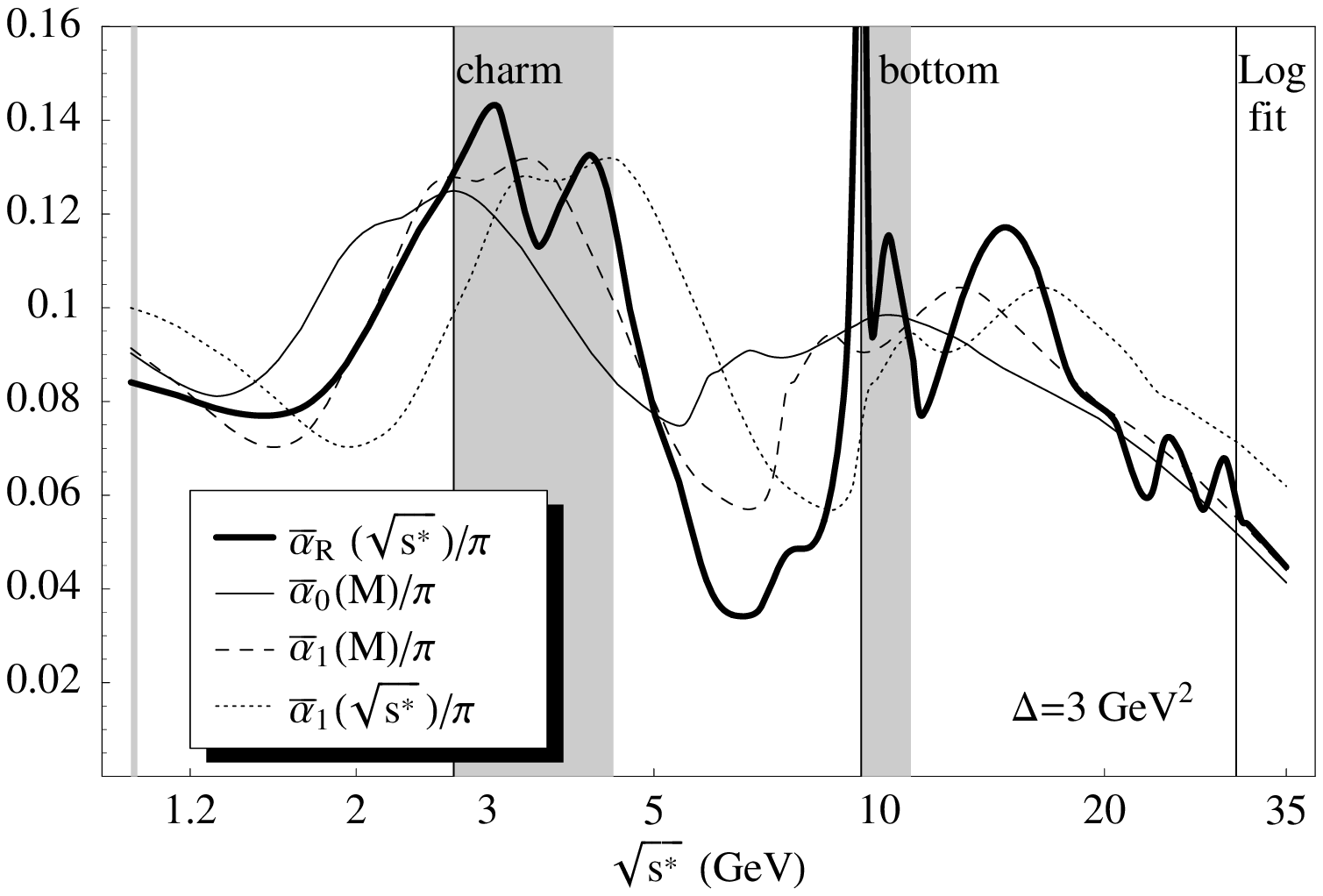,width=8.4cm}
}
{\footnotesize {\bf Figure 3:}
Comparison between $\bar\alpha_R(\sqrt{s^*})$ and different
$\bar{\alpha}_k$ moments at $M=\sqrt{s^*}/\lambda_k$.
The dotted line shows how the agreement is spoilt
if we do not shift $\sqrt{s^*}$ to $M$.
}
\end{figure}

In Fig.3 we compare $\bar{\alpha}_R$ at $\sqrt{s^*}$ with
 $\bar{\alpha}_k$  moments at $M=\sqrt{s^*}/\lambda_k$.
For $\alpha_0$ the agreement is poor, since the low energy region
is not suppressed enough. But for $\alpha_1$
we find a reasonable agreement in several regions, and
we also show
how this agreement disappears if we do not shift the argument
of $\alpha_1$ from $\sqrt{s^*}$ to
$M=\sqrt{s^*}/\lambda_1$.
Starting from higher energies, we find above 30 GeV that
commensurate scale relations are satisfied almost identically, which
is not surprising since above that energy we have fitted with a QCD
inspired behavior. From 15 GeV up to 30 GeV
different experiments have measured rather different central values
at very similar, or even the same, energies. 
The smooth interpolation of these points produces
artificial oscillations around the mean values of the data.
As far as these oscillations are centered
on the $\alpha_k$ curves, there 
is a reasonable agreement, given the quality of the data.
In the region between 5 and 10 GeV 
 there seems to be some controversy about the compatibility
between different experiments (see Fig.1 and 
ref.\cite{discrepancy}). It has become standard not to use
the older data (which is higher) as we have done in Fig.1.
Although the more recent data may be compatible within their
experimental errors with the QCD expectations, their central values
are systematically lower, which is why eq. (\ref{atk}) does not
seem to hold. 
Once there are more accurate data, the tests
we are proposing, together with a thorough error analysis,
will shed light on this situation.

\begin{figure}
\hbox{
\psfig{file=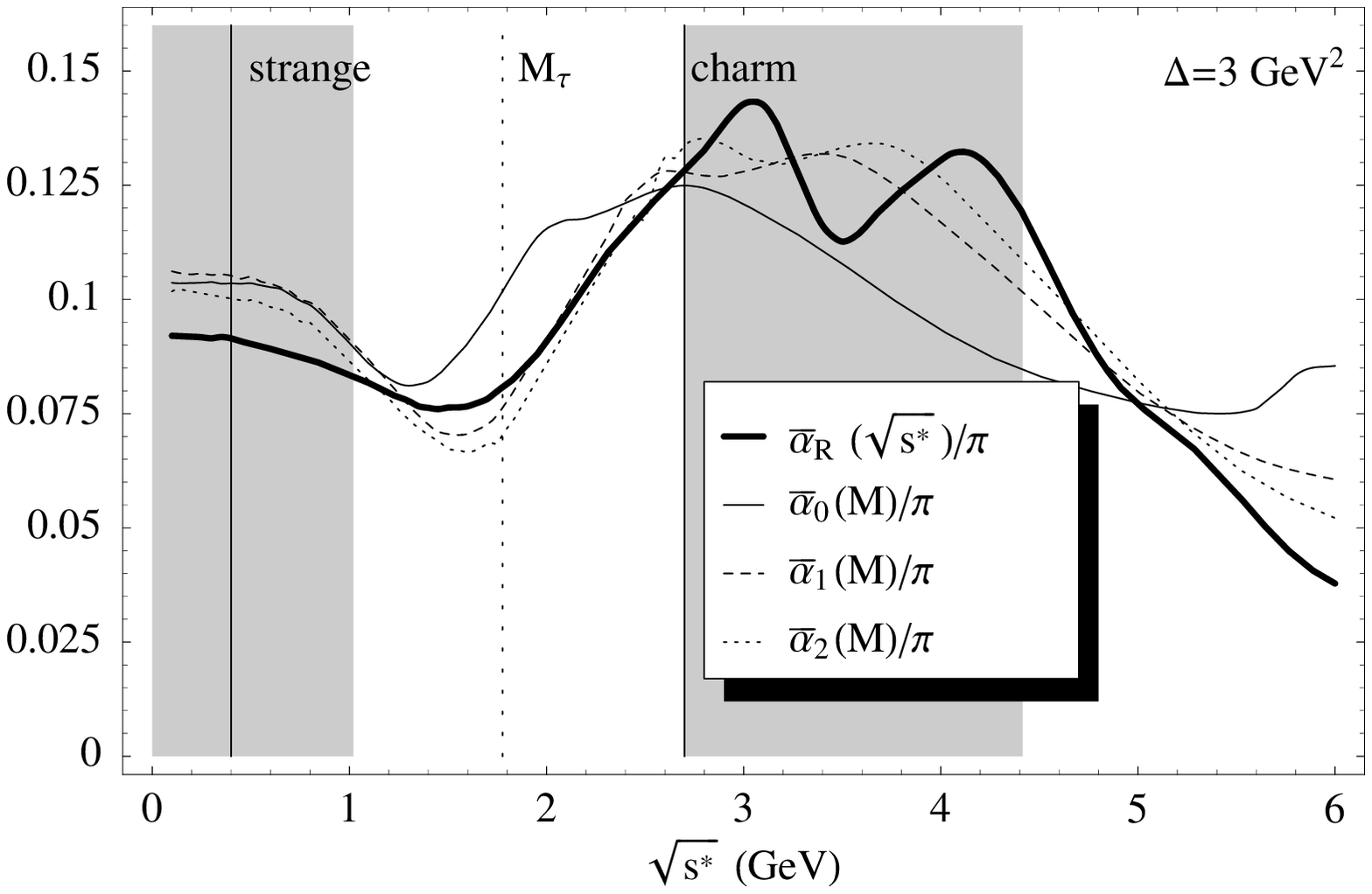,width=8.5cm}
}
{\footnotesize {\bf Figure 4:}
Comparison between $\bar\alpha_R(\sqrt{s^*})$ and different
$\bar{\alpha}_k$ moments at $M=\sqrt{s^*}/\lambda_k$ 
in the low energy region.}
\end{figure}
The low energy region is shown in Fig.4 in more detail. Taking into
account that we are only using LO QCD and central data values,
the agreement between the shaded regions looks quite satisfactory.
This is encouraging for the real $\tau$ lepton, which
sits in a region where PQCD results may be applicable since
it is primarily sensitive to the light $u,d,s$ flavors.
Nevertheless, by looking at energies $\sqrt{s}\sim 1.5$ GeV,
our results seem to support the claims that the $R_{e^+e^-}$
data could be 6-7\%
lower than the QCD expectations in that region
\cite{Groote}.

The commensurate scale relations connecting the moments of the lepton
hadronic decay spectrum to $R_{e^+e^-}$ derived here are basic
scheme-independent tests of PQCD, depending only on the
the universal terms of the $\beta$ function.  We have seen, however,
that a direct comparison with data is problematic
because of several factors such as  the distortions of narrow and broad
resonances, the physical effects
of the quark pair thresholds and the imprecision of much of the
$R_{e^+e^-}$ data. Smearing the data over an energy range
 helps but does not totally remove
the effects due to final-state interactions.  Quark threshold
distortions are partially alleviated by using the Schwinger
corrections at small velocity, but the domain  of  non-relativistic
velocity introduces its own complications, including sensitivity of the
running coupling to the soft
$\alpha_s\, m_q$ scale \cite{Hoang}. Remarkably, the mass range of the
physical $\tau$ lepton is potentially clear of the finite quark mass
effect since it is well below the $c \bar c$ threshold. 
However, it is clear that higher precision
measurements of
$R_{e^+e^-}$ throughout the energy domain below the $Z^0$ boson are needed.

We are indebted to M. Swartz for assistance with
the experimental data. J.R.P. thanks the Spanish 
Ministerio de Educaci\'on y Cultura for financial support.


\begin{thebibliography}{99}

\footnotesize

\bibitem{Braaten}
E. Braaten,  \PRL{60} (1988), \PR{D39} 1458 (1989); 
E. Braaten, S. Narison and A. Pich, \NP{B373} 581 (1992).

\bibitem{Groote} S. Groote {\em et al.}, \PRL{79} 2763 (1997).

\bibitem{Davier} F. Le Diberder and A. Pich, \PL{B289} 165 (1992);
M. Davier, hep-ph/9802447; R. Barate {\em et al}.
Eur. Phys. J. {\bf C4} 409 (1998). K. Ackerstaff {\em et al.}, hep-ex/9808019.

\bibitem{Smilga} B. Chibishov, R.D. Dikeman, 
M. Shifman and N. Uraltsev, Int. Jour. Mod. Phys. {\bf A12} 2075 (1997);  
V.A. Novikov {\em et al.}, \NP{B237} 525 (1984).

\bibitem{Kataev}
S. J. Brodsky,  G. T. Gabadadze, A. L. Kataev and
 H.J. Lu \PL{B372} 133 (1996).

\bibitem{CSR} S. J. Brodsky  and H. J. Lu \PR{D51} 3652 (1995).

\bibitem{effcha}
G. Grunberg, \PR{D29} 2315 (1984).

\bibitem{taudec} S. G. Girishny, A.L. Kataev and S.A. Larin,
\PL{B259} 144 (1991).

\bibitem{PQW} E. C. Poggio, H. R. Quinn and S. Weinberg, \PR{D13} 1958
(1976).

\bibitem{QCDSch}  T. W. Appelquist and H. D. Politzer, \PRL{34}, 43 (1975);
\PR{D12} 1404 (1975).

\bibitem{Sch} J. Schwinger, {\it Particles, Sources and Fields}, Vol.II,
Addison-Wesley, New York, 1973.

\bibitem{MaSt} A. C. Mattingly and P.M. Stevenson, \PR{D49} 437 (1994).

\bibitem{DATA} A. Quenzer , \PL{B76} 512 (1978).
J.Burmeister{\em et al.}, \PL{76B} 361 (1978);
 Ch. Berger {\em et al.}., \PL{81B} 410 (1979);
C. Bacci {\em et al.}, \PL{B86} 234 (1979);
J.L. Siegrist {\em et al.},\PR{D26} 969 (1982);
B.Niczyporuk {\em et al.},\ZP{C15} 299 (1982);
L. M. Barkov {\em et al.}, \NP{B256} 365 (1985);
Z. Jakubobski {\em et al.}, \ZP{C40} 49 (1988);
D. Bisello {\em et al.}, \PL{B220} 325 (1989);
W. Bartel {\em et al.}, \PL{129B} 145 (1983); 337 {\bf 160B} (1985);
B. Naroska {\em et al.}. { Phys. Rep.} {\bf 148} 67 (1987);
B. Adeva {\em et al.}, \PR{D34} 681 (1986);
R. Brandelik {\em et al.}, \PL{113B} 499 (1982);
M. Althoff {\em et al.}, \PL{138B} 441 (1984);
H.-J. Behrend {\em et al.}, \PL{B183} 407 (1987).

\bibitem{discrepancy} M. L. Swartz, \PR{D53} 5268 (1996).

\bibitem{Hoang} V.A. Novikov {\em et al.}, Phys. Rept.{\bf 41} 1 (1978);
M. B. Voloshin, Int. Jour. Mod. Phys. {\bf A10} 2865 (1995);
  S. J. Brodsky, A.H. Hoang, J.H. Kuhn and T. Teubner.
\PL{B359} 355 (1995).

\end{thebibliography}
\end{document}